\documentclass[conference]{IEEEtran}
\IEEEoverridecommandlockouts

\usepackage{cite}
\usepackage{amsmath,amssymb,amsfonts}
\usepackage{graphicx}
\usepackage{textcomp}
\usepackage{xcolor}
\usepackage{algorithm}
\usepackage{algpseudocode}
\usepackage{multirow}
\usepackage{pgfplots}
\pgfplotsset{compat=1.17}
\usepackage{pgfplotstable}
\usepackage{caption}
\usepackage[compatibility=false]{caption}
\pgfplotsset{compat=1.18}
\usepackage{tikz}
\usetikzlibrary{pgfplots.groupplots}
\usepackage{booktabs}
\usepackage{caption}
\usepackage{float}
\usepackage{arydshln}

\pgfplotsset{compat=newest}


\pgfplotsset{
  colormap={RdWhGn}{
    rgb255=(165,15,21);
    rgb255=(222,45,38);
    rgb255=(251,106,74);
    rgb255=(252,174,145);
    rgb255=(254,229,217);
    rgb255=(255,255,255);
    rgb255=(229,245,224);
    rgb255=(199,233,192);
    rgb255=(116,196,118);
    rgb255=(49,163,84);
    rgb255=(0,109,44);
  }
}

\def\BibTeX{{\rm B\kern-.05em{\sc i\kern-.025em b}\kern-.08em
    T\kern-.1667em\lower.7ex\hbox{E}\kern-.125emX}}

\begin{document}

\title{Where to Defend? Layer-Wise Adversarial Training for Robust Transformer-Based Semantic Communications
\thanks{This work is supported by the American University of Beirut University Research Board (URB) and Vertically Integrated Projects (VIP) Program.}}

\author{
\IEEEauthorblockN{Maria Slim, Razane Tajeddine, Mariette Awad, Hadi Sarieddeen}
\IEEEauthorblockA{\textit{Department of Electrical and Computer Engineering} \\
\textit{American University of Beirut}, Beirut 1107 2020, Lebanon \\
mas194@mail.aub.edu; \{razane.tajeddine, mariette.awad, hadi.sarieddeen\}@aub.edu.lb}
}

\maketitle

\begin{abstract}

Deep learning-based semantic communication (DeepSC), a Transformer-based encoder--decoder, achieves semantic fidelity over noisy channels but remains vulnerable to adversarial perturbations injected at multiple stages of the communication pipeline. We present a systematic layer-wise robustness framework that first compares fast gradient sign method (FGSM), projected gradient descent (PGD), and $\ell_2$-normalized fast gradient method (FGM) defenses at the embedding output, and then uses PGD to analyze three attack and defense points: the embedding output, encoder output, and channel-encoder bottleneck. We evaluate text reconstruction on Europarl and UK~Hansard and binary sentiment classification on SST2 and YELP under additive white Gaussian noise (AWGN) and Rayleigh fading. A first-order damage budget, $\varepsilon$ times the $\ell_1$ norm of the clean-input loss gradient at each injection point, predicts the observed attack-severity ordering across the pipeline, and the transfer matrix reveals asymmetric defense transfer: the embedding defense transfers strongly to encoder attacks, whereas encoder defenses generally degrade robustness against upstream attacks. For reconstruction, encoder-point training yields the largest matched gain but fails severely under embedding attacks; joint embedding-plus-encoder training retains comparable gains under encoder attacks while substantially mitigating this mismatch failure, and Rayleigh fading attenuates both robustness gains and degradation. For classification, three of the four defenses collapse to constant predictors; only the channel-encoder defense remains non-degenerate, suggesting a protective role for the $128$D-to-$16$D bottleneck. At a signal-to-noise ratio (SNR) of 9 dB and perturbation budget $\varepsilon=0.3$, the matched encoder defense recovers bilingual evaluation understudy (BLEU) from $\approx 0.10$ to $\approx 0.64$ on Europarl/AWGN, whereas the mismatched encoder defense yields a $-0.444$ BLEU change relative to the undefended baseline under an embedding attack.

\end{abstract}

\section{Introduction}

Semantic communications enable next-generation networks by transmitting meaning rather than raw bits, improving efficiency in bandwidth-limited and noisy environments. Deep learning-based semantic communication (DeepSC)~\cite{xie2021deep}, a Transformer-based encoder--decoder, achieves high task-level fidelity, measured by bilingual evaluation understudy (BLEU) and sentence similarity, over additive white Gaussian noise (AWGN) and fading channels but, like other deep learning-based semantic communication systems, is vulnerable to adversarial semantic noise: small norm-bounded perturbations applied at the source~\cite{hu2022robust} or the embedding representation~\cite{peng2022robust} that corrupt the conveyed semantics. Unlike stochastic channel noise, adversarial perturbations are structured and worst-case~\cite{szegedy2014intriguing}, motivating analysis of how vulnerability and defense effectiveness vary across different stages of the communication pipeline. 
Recent work has also explored architectural defenses against adversarial perturbations, including SignDeepSC, which uses semantic signatures and decoder-side self-repair to improve robustness under FGSM and PGD attacks~\cite{alhaj2026signdeepsc}. Beyond conventional AWGN and Rayleigh settings, DeepSC has also been studied under THz block- and fast-fading channels, extending its evaluation to more challenging channel impairments~\cite{ismail2026thz}.

Adversarial natural language processing (NLP) literature includes white-box attacks (HotFlip~\cite{ebrahimi2018hotflipwhiteboxadversarialexamples}, gradient-guided Transformer attacks~\cite{guo-etal-2021-gradient}) and black-box methods (probability-weighted word saliency~\cite{ren-etal-2019-generating}, TextFooler~\cite{jin2020bertreallyrobuststrong}, genetic synonym substitution~\cite{alzantot2018generatingnaturallanguageadversarial}). Adversarial training via the minimax formulation solved with projected
gradient descent (PGD)~\cite{madry2017towards} dominates among empirical
defenses. Under adaptive attacks, many defenses that combine additional
mechanisms with adversarial training yield lower robustness than
adversarial training alone~\cite{tramer2020adaptive}, while
curriculum-style variants such as friendly adversarial training achieve
robustness without compromising natural
generalization~\cite{zhang2020attackskilltrainingmake}. Robustness efforts in semantic communication include pairing iterative FGSM
with masked Transformers and vector-quantized
codebooks~\cite{hu2022robust,10101778}, calibrated
self-attention~\cite{peng2022robust}, large language model based text
repair~\cite{11112524}, and robust speech
pipelines~\cite{weng2025robustsemanticcommunicationsspeech}.

In practice, semantic communication systems must withstand heterogeneous adversarial conditions where attacks target different architectural locations and interact with channel fading. We propose a systematic robustness analysis framework that compares FGSM-, PGD-, and $\ell_2$-normalized fast gradient method (FGM) based adversarial training at the embedding output, then uses PGD for a layer-wise attack and defense study in a common training and evaluation pipeline. Our contributions are: (i)~layer-wise adversarial vulnerability characterization at embedding, encoder, and channel-encoder points on Europarl and UK~Hansard under AWGN and Rayleigh; (ii)~asymmetric defense-transfer analysis showing that upstream defense transfers strongly from the embedding to the encoder attack point, while defenses placed downstream of the embedding attack degrade robustness most of the time; (iii)~a first-order damage-budget diagnostic based on the $\ell_1$ norm of the clean-input loss gradient at each injection point, which predicts the observed attack-severity ordering before any attack is constructed; and (iv)~a joint embedding-plus-encoder defense that retains comparable gains under encoder attacks while substantially mitigating the encoder-defense mismatch failure, together with a binary-classification stress test (SST2, YELP) exposing task-specific limits of layer-wise adversarial training. We restrict attention to white-box continuous-space attacks (FGSM, PGD, FGM); discrete token-space attacks such as TextFooler and HotFlip target a different threat model and lie outside the scope.

Throughout the paper, non-bold ($a$, $A$), bold lowercase ($\mathbf{a}$), and bold uppercase ($\mathbf{A}$) letters denote scalars, vectors, and matrices, respectively; higher-order tensors are in bold uppercase and their type is clear from context. The Euclidean, $\ell_1$, and $\ell_\infty$ norms of $\mathbf{a}$ are $\|\mathbf{a}\|_2$, $\|\mathbf{a}\|_1$, and $\|\mathbf{a}\|_\infty$. We denote by $\varepsilon$ adversarial perturbation budgets, $\boldsymbol{\delta}$ perturbation vectors, and $\Pi_{\|\cdot\|_\infty\le\varepsilon}$ projection onto the $\ell_\infty$ ball of radius $\varepsilon$.

\section{Problem Formulation and System Model}

We adopt DeepSC as the baseline. We assume a white-box adversary with knowledge of the model architecture and weights, capable of injecting bounded $\ell_\infty$ perturbations at one of three intermediate representations: the embedding output, encoder output, or channel-encoder output. Such access is realistic in two settings: (i)~an insider or compromised-device adversary at the transmitter (e.g., a malicious co-tenant on the edge device hosting the encoder, or a tampered firmware module sitting between pipeline stages), and (ii)~a downloadable model with publicly released weights, which lets an external attacker craft transferable perturbations against any deployment of the same checkpoint. In~(ii), the public weights are used only to craft the perturbation offline; injection still requires transmitter-side access, which we assume is obtained through a compromised software component in the deployed pipeline (as in setting~(i)). The channel itself (AWGN/Rayleigh) is non-adversarial; perturbations are added in the representation space, not over the air. Inputs are batches of token indices $\mathbf{S}\!\in\!\mathbb{Z}^{B\times L}$, where $B\!=\!128$ is the batch size and $L\!=\!40$ the sequence length. The embedding layer maps tokens (with positional encoding) to $\mathbf{E}\!\in\!\mathbb{R}^{B\times L\times d_{\mathrm{model}}}$ with $d_{\mathrm{model}}\!=\!128$. The Transformer encoder produces $\mathbf{M}\!\in\!\mathbb{R}^{B\times L\times 128}$; dense channel-encoder layers compress to $\mathbf{X}\!\in\!\mathbb{R}^{B\times L\times 16}$; the channel produces $\mathbf{Y}$; the channel decoder reconstructs $\hat{\mathbf{M}}\!\in\!\mathbb{R}^{B\times L\times 128}$; the Transformer decoder yields logits $\hat{\mathbf{S}}\!\in\!\mathbb{R}^{B\times L\times V}$, where $V$ is the vocabulary size:
\begin{align*}
\mathbf{E}&=\text{EmbeddingLayer}(\mathbf{S}), &\!\!\!\mathbf{M}&=\text{TransformerEncoder}(\mathbf{E}),\\
\mathbf{X}&=\text{ChannelEncoder}(\mathbf{M}), &\!\!\!\mathbf{Y}&=\text{ChannelLayer}(\mathbf{X}),\\
\hat{\mathbf{M}}&=\text{ChannelDecoder}(\mathbf{Y}), &\!\!\!\hat{\mathbf{S}}&=\text{TransformerDecoder}(\hat{\mathbf{M}}).
\end{align*}
The encoder
output $\mathbf{X}$ is power-normalized to unit average power per real dimension
and reshaped into complex channel symbols. The channel layer models AWGN and Rayleigh fading, over which the received signal is
\begin{equation*}
\mathbf{Y}=h\,\mathbf{X}+\mathbf{N},
\end{equation*}
where $\mathbf{N}$ has i.i.d.\ entries $\mathcal{CN}(0,\sigma_n^2)$ and $h$ is the
fading coefficient: $h=1$ for the AWGN channel, while for the Rayleigh channel
$h\sim\mathcal{CN}(0,1)$ is drawn independently for each transmission block and
held constant across the symbols of the block, i.e., flat block fading. Assuming
perfect channel state information at the receiver, zero-forcing equalization is
applied before decoding,
\begin{equation*}
\hat{\mathbf{Y}}=\mathbf{Y}/h=\mathbf{X}+\mathbf{N}/h,
\end{equation*}
so the semantic decoder observes the transmitted symbols corrupted by
fading-scaled noise.  Training minimizes the cross-entropy (CE) loss, $L_{\mathrm{base}}=L_{\mathrm{CE}}(\mathbf{S},\hat{\mathbf{S}})$, computed between the ground-truth token indices $\mathbf{S}$ and the predicted logits $\hat{\mathbf{S}}$ (averaged over the batch and sequence dimensions).

We adopt a two-stage adversarial training strategy. The first stage applies FGSM, PGD, and $\ell_2$-FGM perturbations at the embedding output (post positional encoding) and compares the three defenses under matched embedding attacks. The second stage uses PGD for layer-wise analysis at three architectural points
along the pipeline: embedding (post positional encoding, $128$D), encoder (after
three Transformer layers, $128$D), and channel-encoder (after compression,
$16$D), enabling defense-transferability evaluation when training and evaluation
attack locations differ. Let
$\mathbf{Z}\in\{\mathbf{E},\mathbf{M},\mathbf{X}\}$ denote the intermediate
representation at the attacked layer, $\mathbf{S}$ the target token sequence,
and $\mathrm{Dec}_\theta(\cdot)$ the downstream subnetwork with parameters
$\theta$ mapping $\mathbf{Z}$ to the output logits
$\hat{\mathbf{S}}=\mathrm{Dec}_\theta(\mathbf{Z})$ (e.g., for
$\mathbf{Z}=\mathbf{M}$, $\mathrm{Dec}_\theta(\cdot)$ includes the channel encoder, channel, channel decoder, and
Transformer decoder). All attacks share three hyperparameters: the $\ell_\infty$
budget $\varepsilon$ (the largest allowed per-coordinate perturbation), the step
size $\alpha$ (the per-coordinate $\ell_\infty$ increment applied at each
iteration), and the iteration count $T$. FGSM applies a single-step perturbation
$\boldsymbol{\Delta}=\varepsilon\,\mathrm{sign}\!\big(\nabla_{\mathbf{Z}}L_{\mathrm{CE}}(\mathbf{S},\hat{\mathbf{S}})\big)$,
giving $\mathbf{Z}_{\mathrm{adv}}=\mathbf{Z}+\boldsymbol{\Delta}$, while PGD
iterates $T$ times, projecting onto the $\varepsilon$-ball via
$\Pi_{\|\cdot\|_\infty\le\varepsilon}$:
\begin{equation}
\begin{aligned}
\boldsymbol{\Delta}^{t+1}
&=\Pi_{\|\boldsymbol{\Delta}\|_\infty\le\varepsilon} 
\Big(\boldsymbol{\Delta}^{t}+\alpha\,\mathrm{sign}\Big(\\
&\qquad \nabla_{\boldsymbol{\Delta}^{t}}
L_{\mathrm{CE}}\big(\mathbf{S},
\mathrm{Dec}_\theta(\mathbf{Z}+\boldsymbol{\Delta}^{t})\big)
\Big)\Big),
\end{aligned}
\label{eq:pgd}
\end{equation}
yielding $\mathbf{Z}_{\mathrm{adv}}=\mathbf{Z}+\boldsymbol{\Delta}^{T}$.
In the layer-wise configuration ($\varepsilon\!=\!0.3$,
$\alpha\!=\!1$, $T\!=\!5$) the step size deliberately exceeds the budget: every
update saturates the sign step and the projection clips each coordinate back to
$\pm\varepsilon$; the iteration applies the full budget at every step rather
than accumulating it gradually. Gradients are backpropagated through the
downstream subnetwork $\mathrm{Dec}_\theta$ (injection point to output); for embedding-level attacks this coincides with the full network
after the embedding layer.
The training objective combines clean and adversarial losses:
$L_{\mathrm{train}}=\eta L_{\mathrm{base}}+\beta
L_{\mathrm{adv}}$, where $\eta,\beta\in[0,1]$ weight the clean and adversarial
terms ($\eta=\beta=0.5$ in all experiments),
$L_{\mathrm{adv}}=L_{\mathrm{CE}}(\mathbf{S},\hat{\mathbf{S}}_{\mathrm{adv}})$,
and $\hat{\mathbf{S}}_{\mathrm{adv}}=\mathrm{Dec}_\theta(\mathbf{Z}_{\mathrm{adv}})$
are the logits produced from the perturbed representation.

\textit{Joint embedding-plus-encoder defense.} A fourth model is hardened at
both representations simultaneously: per batch, two independent PGD
perturbations are generated at the embedding and encoder representations
$\mathbf{E}$ and $\mathbf{M}$, using the same layer-wise
configuration and saturating step size as Eq.~\eqref{eq:pgd}:
\begin{equation}
\boldsymbol{\Delta}_j^{(t+1)} = \Pi_{\|\cdot\|_\infty\le\varepsilon}\!\big[\boldsymbol{\Delta}_j^{(t)} + \alpha\,\mathrm{sign}(\mathbf{G}_j^{(t)})\big],\;j\!\in\!\{\mathrm{emb},\mathrm{enc}\},
\end{equation}
where $\mathbf{Z}_{\mathrm{emb}}\!=\!\mathbf{E}$ and
$\mathbf{Z}_{\mathrm{enc}}\!=\!\mathbf{M}$, and
$\mathbf{G}_j^{(t)}\!=\!\nabla_{\boldsymbol{\Delta}_j^{(t)}}
L_{\mathrm{CE}}\big(\mathbf{S},\,\mathrm{Dec}_\theta(\mathbf{Z}_j\!+\!\boldsymbol{\Delta}_j^{(t)})\big)$
is the loss gradient at the perturbed representation, with
$\mathrm{Dec}_\theta$ the downstream subnetwork from point $j$ as in
Eq.~\eqref{eq:pgd}.
Each perturbed representation propagates through the rest of the pipeline, and
the training objective combines clean and adversarial losses as
$L_{\mathrm{total}}=\tfrac{1}{2}L_{\mathrm{clean}}+\tfrac{1}{4}L_{\mathrm{adv}}^{\mathrm{emb}}+\tfrac{1}{4}L_{\mathrm{adv}}^{\mathrm{enc}}$.
We refer to this model as Emb+Enc.

\section{Methodology}

\subsection{Training and Evaluation Configuration}

The DeepSC backbone~\cite{xie2021deep} consists of three encoder/decoder layers, eight attention heads, $d_{\mathrm{model}}\!=\!128$, feed-forward dimension $d_{\mathrm{ff}}\!=\!512$, and a $16$D channel-encoder bottleneck. Models are trained independently on Europarl (European Parliament proceedings) and UK~Hansard (UK Parliament transcripts), each under AWGN and Rayleigh fading; sentences are tokenized, normalized, and filtered to $4$--$30$ tokens. Adam ($5\!\times\!10^{-4}$, $\beta\!=\!(0.9,0.98)$) trains for $80$ epochs at batch size $128$, with channel SNR sampled from $5$--$10$\,dB. FGSM uses $\varepsilon\!=\!0.3$; PGD uses two configurations (embedding-only experiment in Sec.~\ref{sec:emb_results}: $\varepsilon\!=\!0.3$, $\alpha\!=\!0.1$, $T\!=\!3$; layer-wise experiment in Sec.~\ref{sec:layerwise_results}: $\varepsilon\!=\!0.3$, $\alpha\!=\!1$, $T\!=\!5$); FGM uses $\boldsymbol{\Delta}\!=\!\varepsilon\nabla_{\mathbf{E}}L_{\mathrm{CE}}/\|\nabla_{\mathbf{E}}L_{\mathrm{CE}}\|_2$ with $\varepsilon\!=\!0.3$, following R-DeepSC~\cite{peng2022robust}. Models are tested at SNR $0$--$12$\,dB with $\varepsilon\!\in\!\{0.14,0.3,0.7\}$ in the embedding-level experiment of Sec.~\ref{sec:emb_results}, and at SNR $0$--$18$\,dB with $\varepsilon\!=\!0.3$ in the layer-wise experiments of Secs.~\ref{sec:layerwise_results} and~\ref{sec:asymmetric}. We report BLEU-1 as the primary metric since it is the standard DeepSC reconstruction figure of merit~\cite{xie2021deep} and gives a deterministic, finite-vocabulary score comparable across attack and defense conditions. Layer-wise effectiveness is $\Delta_{\mathrm{BLEU}}=\mathrm{BLEU}_{\mathrm{def}}-\mathrm{BLEU}_{\mathrm{orig}}$, where $\mathrm{BLEU}_{\mathrm{orig}}$ is the undefended model evaluated under the same attack, SNR, and channel condition as the defended model.

\subsection{Mechanistic Diagnostics}

To explain what is consistent with layer-wise training success or failure at each location, we trace
the attack along the transmission chain, from the gradient it exploits to the
loss it realises. All quantities are computed per injection point $j\in\{$embedding, encoder, channel encoder$\}$, with $\mathbf{Z}_j$ the representation at that point; $\ell$ is reserved for norms.

\textit{Damage budget.} For an $\ell_\infty$ attack of budget $\varepsilon$, the maximum first-order (linearized) loss increase that a budget-$\varepsilon$ perturbation injected at point $j$ can achieve is
\begin{equation}
s_j \;=\; \varepsilon\,\bigl\|\nabla_{\mathbf{Z}_j}L_{\mathrm{CE}}\bigr\|_1 ,
\label{eq:budget}
\end{equation}
because $\ell_1$ is the dual norm of $\ell_\infty$: the maximiser of
$\nabla L^\top\boldsymbol\delta$ over $\|\boldsymbol\delta\|_\infty\!\le\!\varepsilon$ is
$\boldsymbol\delta=\varepsilon\,\mathrm{sign}(\nabla L)$. This corresponds to the FGSM update and the sign-gradient direction used by PGD. We evaluate $s_j$ on \emph{clean} inputs, so it is a prediction made before any attack is constructed.

\textit{Channel distortion energy.}
$\text{DE}=\|\mathbf{Y}-\mathbf{X}_{\mathrm{tx}}\|_2^2/\|\mathbf{X}_{\mathrm{tx}}\|_2^2$
characterises the channel rather than the model: $\mathbf{X}_{\mathrm{tx}}$ is power-normalized and the noise variance is set by the SNR; under Rayleigh fading, $\mathbf{Y}-\mathbf{X}_{\mathrm{tx}}$ additionally contains the multiplicative fading term, so DE reflects the combined fading-plus-noise distortion of the implemented channel.

\textit{Diagnostic protocol.} All quantities are computed on the full test set at
SNR\,=\,9\,dB, batch size $64$, shuffling disabled so every model sees identical
batches, using each model's final checkpoint. Adversarial states use
the same attack as evaluation (PGD-$\ell_\infty$, $\varepsilon\!=\!0.3$, $\alpha\!=\!1.0$, $T\!=\!3$, zero initialisation); layer-wise training uses $T=5$ (Sec.~III-A), while the diagnostics use a lighter $T=3$ evaluation attack. Gradients are taken with respect to $\mathbf{Z}_j$.

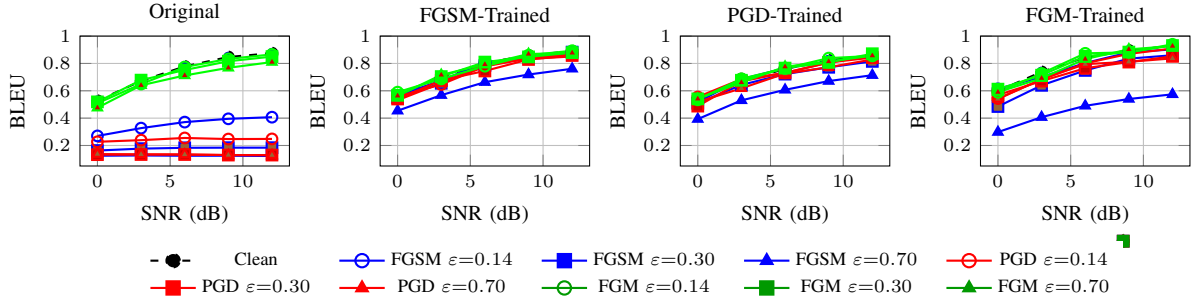
\begin{figure*}[!t]
\centering
\begin{tikzpicture}
\begin{groupplot}[
    group style={group size=4 by 1, horizontal sep=1.2cm},
    width=0.24\linewidth,
    height=3.3cm,
    grid=both,
    ymin=0.05, ymax=1.0,
    xlabel={SNR (dB)}, ylabel={BLEU},
    label style={font=\footnotesize}, tick label style={font=\scriptsize},
    title style={font=\footnotesize, yshift=-1ex}
]
\nextgroupplot[title={Original}]
\addplot+[black, dashed, thick, mark=*] coordinates {(0,0.527) (3,0.672) (6,0.778) (9,0.847) (12,0.875)};
\addplot+[blue, solid, thick, mark=o] coordinates {(0,0.270) (3,0.327) (6,0.371) (9,0.395) (12,0.407)};
\addplot+[blue, solid, thick, mark=square*] coordinates {(0,0.163) (3,0.177) (6,0.183) (9,0.184) (12,0.184)};
\addplot+[blue, solid, thick, mark=triangle*] coordinates {(0,0.125) (3,0.127) (6,0.125) (9,0.125) (12,0.123)};
\addplot+[red, solid, thick, mark=o] coordinates {(0,0.227) (3,0.239) (6,0.255) (9,0.247) (12,0.248)};
\addplot+[red, solid, thick, mark=square*] coordinates {(0,0.133) (3,0.134) (6,0.134) (9,0.130) (12,0.130)};
\addplot+[red, solid, thick, mark=triangle*] coordinates {(0,0.137) (3,0.135) (6,0.134) (9,0.130) (12,0.129)};
\addplot+[green, solid, thick, mark=o] coordinates {(0,0.509) (3,0.649) (6,0.774) (9,0.833) (12,0.866)};
\addplot+[green, solid, thick, mark=square*] coordinates {(0,0.519) (3,0.678) (6,0.752) (9,0.816) (12,0.850)};
\addplot+[green, solid, thick, mark=triangle*] coordinates {(0,0.476) (3,0.640) (6,0.712) (9,0.768) (12,0.810)};

\nextgroupplot[title={FGSM-Trained}]
\addplot+[black, dashed, thick, mark=*] coordinates {(0,0.563) (3,0.701) (6,0.804) (9,0.859) (12,0.890)};
\addplot+[blue, solid, thick, mark=o] coordinates {(0,0.565) (3,0.675) (6,0.807) (9,0.841) (12,0.877)};
\addplot+[blue, solid, thick, mark=square*] coordinates {(0,0.554) (3,0.655) (6,0.784) (9,0.840) (12,0.883)};
\addplot+[blue, solid, thick, mark=triangle*] coordinates {(0,0.454) (3,0.568) (6,0.663) (9,0.719) (12,0.761)};
\addplot+[red, solid, thick, mark=o] coordinates {(0,0.558) (3,0.646) (6,0.778) (9,0.837) (12,0.867)};
\addplot+[red, solid, thick, mark=square*] coordinates {(0,0.540) (3,0.677) (6,0.745) (9,0.829) (12,0.858)};
\addplot+[red, solid, thick, mark=triangle*] coordinates {(0,0.531) (3,0.656) (6,0.795) (9,0.833) (12,0.851)};
\addplot+[green, solid, thick, mark=o] coordinates {(0,0.590) (3,0.687) (6,0.780) (9,0.860) (12,0.893)};
\addplot+[green, solid, thick, mark=square*] coordinates {(0,0.567) (3,0.704) (6,0.810) (9,0.848) (12,0.879)};
\addplot+[green, solid, thick, mark=triangle*] coordinates {(0,0.563) (3,0.716) (6,0.793) (9,0.869) (12,0.877)};

\nextgroupplot[title={PGD-Trained}]
\addplot+[black, dashed, thick, mark=*] coordinates {(0,0.547) (3,0.668) (6,0.760) (9,0.823) (12,0.855)};
\addplot+[blue, solid, thick, mark=o] coordinates {(0,0.518) (3,0.685) (6,0.765) (9,0.815) (12,0.848)};
\addplot+[blue, solid, thick, mark=square*] coordinates {(0,0.509) (3,0.643) (6,0.722) (9,0.772) (12,0.815)};
\addplot+[blue, solid, thick, mark=triangle*] coordinates {(0,0.393) (3,0.530) (6,0.607) (9,0.671) (12,0.714)};
\addplot+[red, solid, thick, mark=o] coordinates {(0,0.555) (3,0.685) (6,0.731) (9,0.806) (12,0.857)};
\addplot+[red, solid, thick, mark=square*] coordinates {(0,0.488) (3,0.669) (6,0.731) (9,0.807) (12,0.843)};
\addplot+[red, solid, thick, mark=triangle*] coordinates {(0,0.519) (3,0.618) (6,0.726) (9,0.772) (12,0.829)};
\addplot+[green, solid, thick, mark=o] coordinates {(0,0.534) (3,0.688) (6,0.764) (9,0.839) (12,0.855)};
\addplot+[green, solid, thick, mark=square*] coordinates {(0,0.541) (3,0.677) (6,0.766) (9,0.808) (12,0.871)};
\addplot+[green, solid, thick, mark=triangle*] coordinates {(0,0.541) (3,0.661) (6,0.776) (9,0.814) (12,0.857)};

\nextgroupplot[title={FGM-Trained}]
\addplot+[black, dashed, thick, mark=*] coordinates {(0,0.599) (3,0.738) (6,0.844) (9,0.895) (12,0.929)};
\addplot+[blue, solid, thick, mark=o] coordinates {(0,0.614) (3,0.701) (6,0.803) (9,0.881) (12,0.903)};
\addplot+[blue, solid, thick, mark=square*] coordinates {(0,0.485) (3,0.640) (6,0.749) (9,0.835) (12,0.860)};
\addplot+[blue, solid, thick, mark=triangle*] coordinates {(0,0.298) (3,0.407) (6,0.490) (9,0.540) (12,0.574)};
\addplot+[red, solid, thick, mark=o] coordinates {(0,0.543) (3,0.683) (6,0.801) (9,0.870) (12,0.909)};
\addplot+[red, solid, thick, mark=square*] coordinates {(0,0.569) (3,0.671) (6,0.765) (9,0.810) (12,0.852)};
\addplot+[red, solid, thick, mark=triangle*] coordinates {(0,0.540) (3,0.685) (6,0.795) (9,0.812) (12,0.837)};
\addplot+[green, solid, thick, mark=o] coordinates {(0,0.589) (3,0.727) (6,0.874) (9,0.881) (12,0.940)};
\addplot+[green, solid, thick, mark=square*] coordinates {(0,0.612) (3,0.714) (6,0.853) (9,0.883) (12,0.931)};
\addplot+[green, solid, thick, mark=triangle*] coordinates {(0,0.590) (3,0.692) (6,0.832) (9,0.903) (12,0.927)};
\end{groupplot}
\node at ($(group c1r1.south)!0.5!(group c4r1.south) + (0,-0.8cm)$) [below] {
\begin{tikzpicture}
\begin{axis}[hide axis, scale only axis, height=0pt, width=0pt, ymin=0, ymax=1,
legend columns=5, legend style={/tikz/every even column/.append style={column sep=0.3cm}, font=\scriptsize, draw=none, nodes={inner sep=2pt}},
legend image post style={mark size=2.5pt, line width=0.8pt}]
\addplot[black, dashed, thick, mark=*] coordinates {(0,0)}; \addlegendentry{Clean}
\addplot[blue, solid, thick, mark=o] coordinates {(0,0)}; \addlegendentry{FGSM $\varepsilon{=}0.14$}
\addplot[blue, solid, thick, mark=square*] coordinates {(0,0)}; \addlegendentry{FGSM $\varepsilon{=}0.30$}
\addplot[blue, solid, thick, mark=triangle*] coordinates {(0,0)}; \addlegendentry{FGSM $\varepsilon{=}0.70$}
\addplot[red, solid, thick, mark=o] coordinates {(0,0)}; \addlegendentry{PGD $\varepsilon{=}0.14$}
\addplot[red, solid, thick, mark=square*] coordinates {(0,0)}; \addlegendentry{PGD $\varepsilon{=}0.30$}
\addplot[red, solid, thick, mark=triangle*] coordinates {(0,0)}; \addlegendentry{PGD $\varepsilon{=}0.70$}
\addplot[green!60!black, solid, thick, mark=o] coordinates {(0,0)}; \addlegendentry{FGM $\varepsilon{=}0.14$}
\addplot[green!60!black, solid, thick, mark=square*] coordinates {(0,0)}; \addlegendentry{FGM $\varepsilon{=}0.30$}
\addplot[green!60!black, solid, thick, mark=triangle*] coordinates {(0,0)}; \addlegendentry{FGM $\varepsilon{=}0.70$}
\end{axis}
\end{tikzpicture}};
\end{tikzpicture}
\caption{BLEU vs.\ SNR for clean and adversarial inputs across FGSM, PGD, and FGM attacks on the original, FGSM-trained, PGD-trained, and FGM-trained DeepSC models (Europarl).}
\label{fig:bleu_vs_snr_all}
\end{figure*}

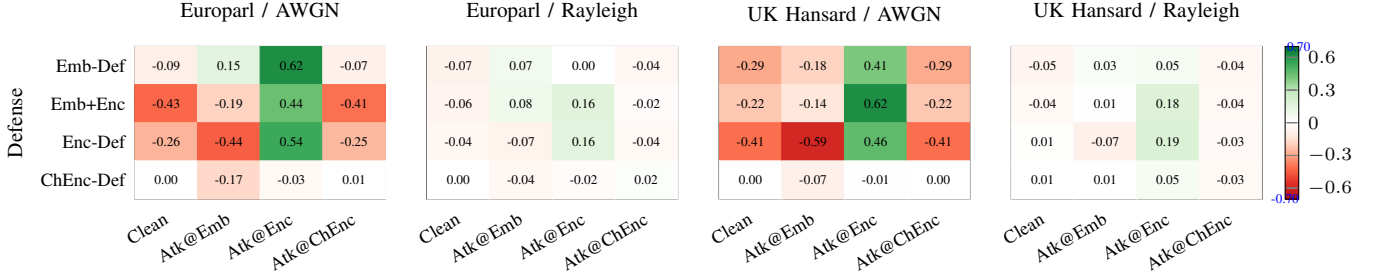
\begin{figure*}[!t]
\centering
\begin{tikzpicture}
\pgfplotsset{
    heatmap style/.style={
        colormap name=RdWhGn,
        point meta min=-0.7, point meta max=0.7,
        width=0.27\linewidth, height=3.6cm,
        enlargelimits=false,
        xtick=data, ytick=data,
        x tick label style={font=\scriptsize, rotate=30, anchor=north east},
        y tick label style={font=\scriptsize},
        title style={font=\footnotesize},
        label style={font=\footnotesize},
        nodes near coords={\pgfmathprintnumber[fixed, fixed zerofill, precision=2]{\pgfplotspointmeta}},
        every node near coord/.append style={font=\tiny, anchor=center, /pgf/number format/assume math mode=true},
        colorbar=false,
    },
}
\begin{groupplot}[
    group style={group size=4 by 1, horizontal sep=0.55cm},
    heatmap style,
]
\nextgroupplot[title={Europarl / AWGN}, ylabel={Defense},
    symbolic x coords={Clean,Atk@Emb,Atk@Enc,Atk@ChEnc},
    symbolic y coords={ChEnc-Def,Enc-Def,Emb+Enc,Emb-Def}]
\addplot[matrix plot*, mesh/cols=4, point meta=explicit] coordinates {
(Clean,Emb-Def)    [-0.086] (Atk@Emb,Emb-Def)    [ 0.149] (Atk@Enc,Emb-Def)    [ 0.615] (Atk@ChEnc,Emb-Def)    [-0.073]
(Clean,Emb+Enc)    [-0.425] (Atk@Emb,Emb+Enc)    [-0.185] (Atk@Enc,Emb+Enc)    [ 0.440] (Atk@ChEnc,Emb+Enc)    [-0.407]
(Clean,Enc-Def)    [-0.263] (Atk@Emb,Enc-Def)    [-0.444] (Atk@Enc,Enc-Def)    [ 0.537] (Atk@ChEnc,Enc-Def)    [-0.251]
(Clean,ChEnc-Def)  [-0.001] (Atk@Emb,ChEnc-Def)  [-0.173] (Atk@Enc,ChEnc-Def)  [-0.033] (Atk@ChEnc,ChEnc-Def)  [ 0.007]
};

\nextgroupplot[title={Europarl / Rayleigh}, ylabel={},
    symbolic x coords={Clean,Atk@Emb,Atk@Enc,Atk@ChEnc},
    symbolic y coords={ChEnc-Def,Enc-Def,Emb+Enc,Emb-Def},
    yticklabels={,,}]
\addplot[matrix plot*, mesh/cols=4, point meta=explicit] coordinates {
(Clean,Emb-Def)    [-0.071] (Atk@Emb,Emb-Def)    [ 0.069] (Atk@Enc,Emb-Def)    [ 0.001] (Atk@ChEnc,Emb-Def)    [-0.044]
(Clean,Emb+Enc)    [-0.059] (Atk@Emb,Emb+Enc)    [ 0.084] (Atk@Enc,Emb+Enc)    [ 0.161] (Atk@ChEnc,Emb+Enc)    [-0.019]
(Clean,Enc-Def)    [-0.044] (Atk@Emb,Enc-Def)    [-0.066] (Atk@Enc,Enc-Def)    [ 0.155] (Atk@ChEnc,Enc-Def)    [-0.037]
(Clean,ChEnc-Def)  [-0.001] (Atk@Emb,ChEnc-Def)  [-0.042] (Atk@Enc,ChEnc-Def)  [-0.019] (Atk@ChEnc,ChEnc-Def)  [ 0.021]
};

\nextgroupplot[title={UK Hansard / AWGN}, ylabel={},
    symbolic x coords={Clean,Atk@Emb,Atk@Enc,Atk@ChEnc},
    symbolic y coords={ChEnc-Def,Enc-Def,Emb+Enc,Emb-Def},
    yticklabels={,,}]
\addplot[matrix plot*, mesh/cols=4, point meta=explicit] coordinates {
(Clean,Emb-Def)    [-0.290] (Atk@Emb,Emb-Def)    [-0.184] (Atk@Enc,Emb-Def)    [ 0.409] (Atk@ChEnc,Emb-Def)    [-0.285]
(Clean,Emb+Enc)    [-0.217] (Atk@Emb,Emb+Enc)    [-0.135] (Atk@Enc,Emb+Enc)    [ 0.616] (Atk@ChEnc,Emb+Enc)    [-0.215]
(Clean,Enc-Def)    [-0.411] (Atk@Emb,Enc-Def)    [-0.587] (Atk@Enc,Enc-Def)    [ 0.463] (Atk@ChEnc,Enc-Def)    [-0.409]
(Clean,ChEnc-Def)  [ 0.000] (Atk@Emb,ChEnc-Def)  [-0.073] (Atk@Enc,ChEnc-Def)  [-0.005] (Atk@ChEnc,ChEnc-Def)  [ 0.004]
};

\nextgroupplot[title={UK Hansard / Rayleigh}, ylabel={},
    symbolic x coords={Clean,Atk@Emb,Atk@Enc,Atk@ChEnc},
    symbolic y coords={ChEnc-Def,Enc-Def,Emb+Enc,Emb-Def},
    yticklabels={,,},
    colorbar, colorbar style={width=0.18cm, font=\scriptsize, ytick={-0.6,-0.3,0,0.3,0.6}}]
\addplot[matrix plot*, mesh/cols=4, point meta=explicit] coordinates {
(Clean,Emb-Def)    [-0.054] (Atk@Emb,Emb-Def)    [ 0.030] (Atk@Enc,Emb-Def)    [ 0.050] (Atk@ChEnc,Emb-Def)    [-0.041]
(Clean,Emb+Enc)    [-0.036] (Atk@Emb,Emb+Enc)    [ 0.010] (Atk@Enc,Emb+Enc)    [ 0.179] (Atk@ChEnc,Emb+Enc)    [-0.036]
(Clean,Enc-Def)    [ 0.008] (Atk@Emb,Enc-Def)    [-0.074] (Atk@Enc,Enc-Def)    [ 0.190] (Atk@ChEnc,Enc-Def)    [-0.030]
(Clean,ChEnc-Def)  [ 0.007] (Atk@Emb,ChEnc-Def)  [ 0.006] (Atk@Enc,ChEnc-Def)  [ 0.048] (Atk@ChEnc,ChEnc-Def)  [-0.034]
};
\end{groupplot}
\end{tikzpicture}
\caption{BLEU improvement vs.\ undefended baseline at SNR\,=\,9\,dB and $\varepsilon=0.3$ (PGD, $\ell_\infty$), as a heatmap. Rows are defenses, columns are attack conditions, panels are dataset/channel combinations. \textcolor{green!50!black}{Green cells} indicate positive improvement (gain), \textcolor{red!70!black}{red cells} indicate degradation, white indicates no change; color saturation encodes magnitude on a symmetric scale. }
\label{fig:bleu_improvement}
\end{figure*}

\section{Results and Discussion}

\subsection{Embedding-Level Robustness}
\label{sec:emb_results}
Fig.~\ref{fig:bleu_vs_snr_all} compares BLEU performance of clean-trained, FGSM-, PGD-, and FGM-trained models across perturbation magnitudes and SNRs. The clean-trained model is highly vulnerable, with sharp degradation under FGSM and PGD. Adversarial training restores robustness across all three methods. FGM achieves competitive robustness at moderate budgets but sometimes degrades under stronger attacks, consistent with the mismatch between $\ell_2$-based training and $\ell_\infty$-based evaluation. Given PGD's strong, consistent robustness and widespread adoption, we use PGD for all subsequent layer-wise analysis.

\begin{figure}[!t]
\centering
\begin{tikzpicture}
\begin{groupplot}[
  group style={group size=2 by 2, horizontal sep=0.95cm, vertical sep=0.45cm},
  width=0.46\linewidth, height=2.5cm, ybar=6pt,
  symbolic x coords={Emb,Enc,ChEnc}, xtick=data, enlarge x limits=0.25,
  tick label style={font=\scriptsize}, label style={font=\footnotesize},
  title style={font=\footnotesize, yshift=-1ex},
]
\nextgroupplot[title={Europarl}, ymode=log, ylabel={$s_j$}, xticklabels={,,}]
\addplot[fill=blue!60, draw=blue!30!black] coordinates {(Emb,0.001698) (Enc,0.003652) (ChEnc,0.000401)};
\nextgroupplot[title={UK Hansard}, ymode=log, xticklabels={,,}]
\addplot[fill=blue!60, draw=blue!30!black] coordinates {(Emb,0.001398) (Enc,0.003072) (ChEnc,0.00031)};
\nextgroupplot[ylabel={BLEU degr.\ (\%)}, ymin=0, ymax=100]
\addplot[fill=red!60, draw=red!30!black] coordinates {(Emb,25.73) (Enc,88.94) (ChEnc,1.555)};
\nextgroupplot[ ymin=0, ymax=100]
\addplot[fill=red!60, draw=red!30!black] coordinates {(Emb,22.72) (Enc,89.15) (ChEnc,0.9804)};
\end{groupplot}
\end{tikzpicture}
\caption{Undefended baseline: first-order damage budget $s_j=\varepsilon\|\nabla_{\mathbf{Z}_j}L\|_1$ of Eq.~\eqref{eq:budget} at each injection point, computed on clean inputs (top, log scale), against the BLEU degradation measured when the PGD attack is run at that point (bottom). SNR\,=\,9\,dB, $\varepsilon=0.3$, seed 242.}
\label{fig:baseline_budget}
\end{figure}
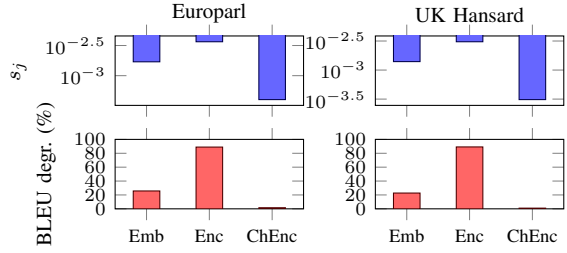

\subsection{Layer-Wise Vulnerability and Defense Transfer}
\label{sec:layerwise_results}

We evaluate the three single-point PGD defenses (Emb-Def, Enc-Def, ChEnc-Def) and the joint Emb+Enc defense against PGD attacks at the same three points, on Europarl and UK~Hansard, AWGN and Rayleigh, at SNR\,=\,9\,dB and $\varepsilon\!=\!0.3$. Fig.~\ref{fig:bleu_improvement} reports BLEU improvement relative to the undefended baseline, decomposing each defense outcome into its clean-accuracy cost and its robustness gain.

Attack severity tracks the first-order damage budget. On the undefended
Europarl/AWGN model, the clean-input budget
$\varepsilon\|\nabla_{\mathbf{Z}_j}L_{\mathrm{CE}}\|_1$ ranks the three
injection points in the same order as adversarial BLEU: the encoder has the
largest budget and suffers the greatest BLEU collapse, the embedding is
intermediate, and the channel encoder has the smallest budget and highest
attacked BLEU.
Thus, the clean-input gradient predicts where the attack damages most, and the
budget is computed from clean inputs with no attack constructed.
Unlike the remaining figures, which average seeds $42$ and $142$, the
damage-budget results here use an independently trained seed ($242$); the
adversarial-BLEU orderings agree across the two sets.
The $16$D channel-encoder bottleneck constrains attacks at that point: the
adversary operates in a $16$D representation rather than $128$D
and the bottleneck-output gradient is the smallest of the three. As reported in
Fig.~\ref{fig:bleu_improvement}, matched defenses recover BLEU at a fixed clean
cost set by defense location: encoder defense lifts BLEU from $\approx 0.10$ to
$\approx 0.64$ at clean cost $-0.26$ on Europarl/AWGN; embedding defense lifts
BLEU from $\approx 0.71$ to $\approx 0.85$ at clean cost $-0.086$;
channel-encoder defense yields negligible matched gain at near-zero clean cost.
A crossover analysis across the full SNR $0$--$18$\,dB range confirms the matched defense remains the most reliable choice for both datasets, occasionally overtaken by upstream transferred defenses on AWGN.

\subsection{Asymmetric Defense Transfer and Pronounced Mismatch}
\label{sec:asymmetric}

The transfer matrix in Fig.~\ref{fig:bleu_improvement} is directionally asymmetric: the embedding defense transfers strongly downstream to the encoder attack, while the encoder defense generally degrades robustness under the upstream embedding attack; no defense meaningfully improves over the baseline under the channel-encoder attack, whose damage on the undefended model is already negligible. On Europarl/AWGN, embedding defense improves BLEU by $+0.615$ under encoder attack, exceeding the matched encoder defense ($+0.537$); encoder defense against an embedding attack instead incurs $-0.444$ on Europarl and $-0.587$ on UK~Hansard.

\begin{table*}[!t]
\centering
\caption{Headline findings, evidence, and scope conditions.}
\label{tab:findings}
\renewcommand{\arraystretch}{1.4}\setlength{\tabcolsep}{5pt}\footnotesize
\begin{tabular}{@{}p{0.18\textwidth} p{0.40\textwidth} p{0.30\textwidth}@{}}
\toprule
\textbf{Claim} & \textbf{Evidence} & \textbf{Caveat / scope} \\
\midrule
Encoder attacks are the most damaging single-point attacks. & Undefended BLEU drops to $\approx0.10$ under attack at the encoder vs.\ $\approx0.71$ under attack at the embedding and $\approx0.96$ under attack at the channel-encoder; the damage budget $\varepsilon\|\nabla_{\mathbf{Z}_j}L\|_1$ at the encoder is the largest of the three points. & Holds for both Europarl and UK~Hansard under AWGN at SNR\,=\,9\,dB; ordering preserved but compressed under Rayleigh. \\
\addlinespace[3pt]\hdashline\addlinespace[3pt]
Embedding defense transfers strongly to the encoder attack. & Embedding defense gives $+0.149/+0.615/-0.073$ under attack at the embedding/encoder/channel-encoder on Europarl/AWGN; no undefended prefix exists upstream of the embedding point. & Transfer benefit is largest under AWGN; under Rayleigh all transfer effects compress toward zero. \\
\addlinespace[3pt]\hdashline\addlinespace[3pt]
Encoder defense has strong matched gain but poor upstream transfer. & Encoder defense: $+0.537$ at matched attack at the encoder, but $-0.444/-0.587$ under attack at the embedding on Europarl/UK~Hansard AWGN. & The mismatch failure appears compressed under Rayleigh ($-0.07$). \\
\addlinespace[3pt]\hdashline\addlinespace[3pt]
Channel-encoder bottleneck limits attack headroom. & Attack at the channel-encoder reduces undefended BLEU only by $\approx0.04$; channel-encoder defense matched gain $\le 0.021$; the damage budget at this point is the smallest of the three, indicating limited attack and defense headroom at the $16$D channel-encoder representation.
 & Specific to the $128\!\to\!16$ ratio in DeepSC \\
\addlinespace[3pt]\hdashline\addlinespace[3pt]
Rayleigh compresses both gains and damages. & Europarl AWGN $\to$ Rayleigh: matched encoder-defense gain $+0.537\!\to\!+0.155$, mismatched failure $-0.444\!\to\!-0.066$; baseline channel distortion higher and damage budgets smaller than under AWGN. &

Compression may reflect weakened attack optimization under per-step channel resampling. In our setting, AWGN-only validation overestimates both defense value and risk. \\
\addlinespace[3pt]\hdashline\addlinespace[3pt]
Classification behaves differently from reconstruction. & On SST2 and YELP, three of four defenses reach
accuracy near $1-p$, where $p$ is the majority-class fraction,
and Fisher $F<10^{-3}$; only channel-encoder defense remains
non-degenerate. & Specific to $K\!=\!2$ at $\varepsilon\!=\!0.3$; we conjecture $K\!\ge\!4$ or $\varepsilon\!\le\!0.1$ would recover non-trivial behavior. \\
\bottomrule
\end{tabular}
\end{table*}

\subsection{Stress-Test: Generalization to Classification}
\label{sec:classification}

We replace the Transformer decoder with a classification head with $K$ output classes and retrain
end-to-end on SST2 and YELP ($K{=}2$, $\approx781$ test sentences each) across
two independent seeds. Accuracy is reported throughout as the fraction
of correctly classified test sentences, so a two-class problem has a chance
level of approximately $0.50$. Alongside accuracy we measure the encoder-output
Fisher ratio
\begin{equation}
F=\frac{\|\boldsymbol{\mu}_+-\boldsymbol{\mu}_-\|^2}
{\operatorname{tr}\boldsymbol{\Sigma}_++\operatorname{tr}\boldsymbol{\Sigma}_-},
\label{eq:fisher}
\end{equation}
where $\boldsymbol{\mu}_c$ and $\boldsymbol{\Sigma}_c$ are the mean and covariance
of the masked mean-pooled encoder output over the test sentences of
sentiment class $c\in\{+,-\}$, corresponding to positive and
negative sentiment, and both traces sum over the
$d_{\mathrm{model}}=128$ latent coordinates. In words, $F$ asks whether
sentences of the two classes come to rest in different regions of the encoder
representation: a large $F$ means the two class clouds are far apart relative
to their spread; $F\!\to\!0$ indicates vanishing mean
separation relative to within-class spread, not implying coincident class
distributions or absence of nonlinear separability.

Table~\ref{tab:cls_collapse} shows that three of the four defenses do not survive
their own training. Before any attack is applied, Emb-Def, Enc-Def and Emb+Enc
sit within $0.01$ accuracy of $1-p$, where $p$ is the majority-class fraction of
the test set ($p=0.510$ for SST2 and $p=0.504$ for YELP), which is the accuracy
obtained by always predicting the minority class. These networks
have not merely become less accurate; they have stopped discriminating. The Fisher
ratio, computed independently of the classifier output, confirms the same conclusion
from inside the representation: $F$ falls below $10^{-3}$ against $1.6$ to $3.4$
for the working models, indicating near-zero mean separation between
positive and negative sentences relative to their within-class variance. Only
ChEnc-Def escapes, retaining both accuracy ($0.838 $ to $0.889$) and separability.

Surviving training does not imply robustness. Evaluating every model
against PGD injected at each of the three points shows first that the attack
surface is strongly asymmetric: perturbations at the encoder or channel-encoder
output leave every model close to its clean accuracy, and essentially all damage
comes from perturbing the token embeddings. Under that embedding attack
ChEnc-Def, the only defense that is still a functioning classifier, performs
worse than the undefended baseline in every configuration, losing a further
$0.05$ to $0.11$ accuracy on SST2 and approximately $0.31$ on YELP. The larger
vulnerability occurs on the longer-sequence YELP dataset, where sentences are longer on average ($21.4$ against $15.0$ tokens): a fixed per-coordinate
$\varepsilon$ could be granting a longer input a proportionally larger total
perturbation.

\begin{table}[t]
\centering
\caption{Classification collapse confirmation. Clean accuracy at SNR\,=\,9\,dB
and the encoder-output Fisher ratio $F$ of Eq.~\eqref{eq:fisher}, the squared
distance between the class-conditional means of the pooled encoder output
normalised by the summed within-class variances. Accuracy is the fraction of
correctly classified test sentences. Bold marks jointly collapsed cells:
accuracy within $0.01$ of $1-p$, the minority-class fraction, \emph{and} $F<10^{-3}$.}
\label{tab:cls_collapse}
\renewcommand{\arraystretch}{1.0}\setlength{\tabcolsep}{2pt}\footnotesize
\begin{tabular}{@{}llccccc@{}}
\toprule
Dataset & Channel & Orig & Emb-Def & Enc-Def & ChEnc-Def & Emb+Enc \\
\midrule
\multirow{4}{*}{SST2} & AWGN acc & $0.834$ & $\mathbf{0.490}$ & $\mathbf{0.490}$ & $0.838$ & $\mathbf{0.500}$ \\
 & \quad $F$ & $1.81$ & $\mathbf{0.00}$ & $\mathbf{0.00}$ & $1.65$ & $\mathbf{0.00}$ \\
 & Rayleigh acc & $0.826$ & $\mathbf{0.490}$ & $\mathbf{0.490}$ & $0.848$ & $\mathbf{0.490}$ \\
 & \quad $F$ & $1.60$ & $\mathbf{0.00}$ & $\mathbf{0.00}$ & $2.07$ & $\mathbf{0.00}$ \\
\midrule
\multirow{4}{*}{YELP} & AWGN acc & $0.876$ & $\mathbf{0.496}$ & $\mathbf{0.496}$ & $0.875$ & $\mathbf{0.496}$ \\
 & \quad $F$ & $2.97$ & $\mathbf{0.00}$ & $\mathbf{0.00}$ & $2.97$ & $\mathbf{0.00}$ \\
 & Rayleigh acc & $0.879$ & $\mathbf{0.496}$ & $\mathbf{0.496}$ & $0.889$ & $\mathbf{0.496}$ \\
 & \quad $F$ & $3.12$ & $\mathbf{0.00}$ & $\mathbf{0.00}$ & $3.43$ & $\mathbf{0.00}$ \\
\bottomrule
\end{tabular}
\end{table}

\subsection{Findings Summary}
\label{sec:takeaways}

Table~\ref{tab:findings} consolidates the six headline findings of the layer-wise study with the evidence supporting each and the conditions under which it holds, intended both as a reference for navigating the results above and as a precise statement of scope for follow-up work.

\section{Limitations}
\label{sec:limitations}

Two caveats bound the generality of the above findings. First, all results are
for a single DeepSC backbone (three-layer encoder/decoder,
$d_{\mathrm{model}}\!=\!128$, $128$D-to-$16$D bottleneck); the
observed protection at the bottleneck in particular is expected to vary with
the compression ratio. Second, we evaluate continuous-space white-box
$\ell_\infty$ attacks only; discrete token-level attacks (HotFlip, TextFooler),
$\ell_2$ and $\ell_0$ threats, and black-box transfer attacks target a different threat model and lie outside the scope.

\section{Conclusions}

We presented a systematic robustness analysis framework for DeepSC that first
compares FGSM-, PGD-, and FGM-based adversarial training at the embedding
output and then conducts a PGD-only layer-wise study, evaluated on text
reconstruction (Europarl, UK~Hansard) and binary sentiment classification
(SST2, YELP) under AWGN and Rayleigh. Layer-wise analysis decomposes every defense outcome into a fixed
clean cost set by defense location and a robustness gain
empirically limited at the $16$D bottleneck, and shows that
defense transfer is directionally asymmetric: the embedding defense transfers
strongly to encoder attacks, whereas encoder-point
training under AWGN fails against embedding attacks, a failure the
joint Emb+Enc defense substantially mitigates. The first-order damage budget of
Eq.~\eqref{eq:budget}, computed on clean inputs, predicts the observed attack-severity
ordering across injection points, and Rayleigh fading compresses both gains and
damages. The binary-classification evaluation reveals task-specific limits:
three of four defenses collapse to constant predictors during training, and
only ChEnc-Def remains non-degenerate, showing that defense at
the channel-encoder bottleneck (a $128$D-to-$16$D dense projection that
compresses semantic features for transmission) provides the most consistent
cross-task protection in our experiments, though one that still degrades under
embedding attacks. Future work can extend the analysis to other channels
and architectures, explore hybrid multi-point curricula, test
amplification-based explanations of the transfer asymmetry, and probe
classification at lower $\varepsilon$ or $K\!\ge\!4$.

\bibliographystyle{IEEEtran}
\bibliography{references}

@article{xie2021deep,
 author={Xie, Huiqiang and Qin, Zhijin and Li, Geoffrey Ye and Juang, Biing-Hwang},
  journal={IEEE Transactions on Signal Processing}, 
  title={Deep Learning Enabled Semantic Communication Systems}, 
  year={2021},
  volume={69},
  number={},
  pages={2663-2675},
  doi={10.1109/TSP.2021.3071210}
}

@inproceedings{peng2022robust,
  title={A robust deep learning enabled semantic communication system for text},
  author={Peng, Xiang and Qin, Zhijin and Huang, Danlan and Tao, Xiaoming and Lu, Jianhua and Liu, Guangyi and Pan, Chengkang},
  booktitle={GLOBECOM 2022-2022 IEEE Global Communications Conference},
  pages={2704--2709},
  year={2022},
  organization={IEEE}
}

@inproceedings{hu2022robust,
  title={Robust semantic communications against semantic noise},
  author={Hu, Qiyu and Zhang, Guangyi and Qin, Zhijin and Cai, Yunlong and Yu, Guanding and Li, Geoffrey Ye},
  booktitle={2022 IEEE 96th Vehicular Technology Conference (VTC2022-Fall)},
  pages={1--6},
  year={2022},
  organization={IEEE}
}

@inproceedings{madry2017towards,
  title={Towards deep learning models resistant to adversarial attacks},
  author={Madry, Aleksander and Makelov, Aleksandar and Schmidt, Ludwig and Tsipras, Dimitris and Vladu, Adrian},
  booktitle={International Conference on Learning Representations (ICLR)},
  year={2018}
}

@article{szegedy2014intriguing,
  title={Intriguing properties of neural networks},
  author={Szegedy, Christian and Zaremba, Wojciech and Sutskever, Ilya and Bruna, Joan and Erhan, Dumitru and Goodfellow, Ian and Fergus, Rob},
  journal={arXiv preprint arXiv:1312.6199},
  year={2013}
}

@article{tramer2020adaptive,
  title={On adaptive attacks to adversarial example defenses},
  author={Tramer, Florian and Carlini, Nicholas and Brendel, Wieland and Madry, Aleksander},
  journal={Advances in neural information processing systems},
  volume={33},
  pages={1633--1645},
  year={2020}
}

@inproceedings{jin2020bertreallyrobuststrong,
  title={Is {BERT} really robust? a strong baseline for natural language attack on text classification and entailment},
  author={Jin, Di and Jin, Zhijing and Zhou, Joey Tianyi and Szolovits, Peter},
  booktitle={Proceedings of the AAAI conference on artificial intelligence},
  volume={34},
  number={05},
  pages={8018--8025},
  year={2020}
}

@article{alhaj2026signdeepsc,
  author  = {Khalil Alhaj and Razane Tajeddine and Hadi Sarieddeen},
  title   = {{SignDeepSC}: A Semantic Signature-based Approach for Robust Semantic Communication},
  journal = {arXiv preprint arXiv:2607.25676},
  year    = {2026}
}

@article{ismail2026thz,
  author  = {Fatima Ismail and Hadi Sarieddeen and Jihad Fahs},
  title   = {Semantic Communications in the {THz} Band},
  journal = {arXiv preprint arXiv:2607.07455},
  year    = {2026}
}

@inproceedings{ebrahimi2018hotflipwhiteboxadversarialexamples,
  title={Hotflip: {W}hite-box adversarial examples for text classification},
  author={Ebrahimi, Javid and Rao, Anyi and Lowd, Daniel and Dou, Dejing},
  booktitle={Proceedings of the 56th Annual Meeting of the Association for Computational Linguistics (Volume 2: Short Papers)},
  pages={31--36},
  year={2018}
}

@inproceedings{ren-etal-2019-generating,
  title={Generating natural language adversarial examples through probability weighted word saliency},
  author={Ren, Shuhuai and Deng, Yihe and He, Kun and Che, Wanxiang},
  booktitle={Proceedings of the 57th annual meeting of the association for computational linguistics},
  pages={1085--1097},
  year={2019}
}

@ARTICLE{11112524,
  author={Park, Sojeong and Noh, Hyeonho and Yang, Hyun Jong},
  journal={IEEE Transactions on Vehicular Technology}, 
  title={Robust Transmission of Punctured Text with Large Language Model-based Recovery}, 
  year={2025},
  volume={},
  number={},
  pages={1-6},
  doi={10.1109/TVT.2025.3595593}}

@inproceedings{weng2025robustsemanticcommunicationsspeech,
  title={Robust semantic communications for speech transmission},
  author={Weng, Zhenzi and Qin, Zhijin and Li, Geoffrey Ye},
  booktitle={ICASSP 2025-2025 IEEE International Conference on Acoustics, Speech and Signal Processing (ICASSP)},
  pages={1--5},
  year={2025},
  organization={IEEE}
}

@ARTICLE{10101778,
  author={Hu, Qiyu and Zhang, Guangyi and Qin, Zhijin and Cai, Yunlong and Yu, Guanding and Li, Geoffrey Ye},
  journal={IEEE Transactions on Wireless Communications}, 
  title={Robust Semantic Communications With Masked {VQ-VAE} Enabled Codebook}, 
  year={2023},
  volume={22},
  number={12},
  pages={8707-8722},
  doi={10.1109/TWC.2023.3265201}}

@inproceedings{guo-etal-2021-gradient,
  title={Gradient-based adversarial attacks against text transformers},
  author={Guo, Chuan and Sablayrolles, Alexandre and J{\'e}gou, Herv{\'e} and Kiela, Douwe},
  booktitle={Proceedings of the 2021 Conference on Empirical Methods in Natural Language Processing},
  pages={5747--5757},
  year={2021}
}

@inproceedings{zhang2020attackskilltrainingmake,
  title={Attacks which do not kill training make adversarial learning stronger},
  author={Zhang, Jingfeng and Xu, Xilie and Han, Bo and Niu, Gang and Cui, Lizhen and Sugiyama, Masashi and Kankanhalli, Mohan},
  booktitle={International conference on machine learning},
  pages={11278--11287},
  year={2020},
  organization={PMLR}
}

@inproceedings{alzantot2018generatingnaturallanguageadversarial,
  title={Generating natural language adversarial examples},
  author={Alzantot, Moustafa and Sharma, Yash and Elgohary, Ahmed and Ho, Bo-Jhang and Srivastava, Mani and Chang, Kai-Wei},
  booktitle={Proceedings of the 2018 conference on empirical methods in natural language processing},
  pages={2890--2896},
  year={2018}
}

\end{document}